\documentstyle[preprint,floats,prl,aps,epsf]{revtex}
\begin{document}
\def\calN{{\cal N}}
\def\beq{\begin{equation}}
\def\eeq{\end{equation}}
\def\ee{\nonumber\end{equation}}
\def\bea{\begin{eqnarray}}
\def\eea{\end{eqnarray}}
\def\frak#1#2{{\textstyle{{#1}\over{#2}}}}
\def\nn{\nonumber\\}
\def\pa{\partial}
\def\mhat{\hat{m}}
\def\Tr{\, {\rm Tr}}
\def\tr{\, {\rm tr}} 
\def\Ytilde{{\tilde Y}}
\def\lf{{16\pi^2\epsilon}}
\def\half{\frac{1}{2}}
\preprint{\vbox{\hbox {December 2004} }}

\draft
\title{Chiral Fermions and Quadratic Divergences} 
\author{\bf Xavier Calmet, Paul H. Frampton and Ryan M. Rohm}
\address{University of North Carolina, Chapel Hill, NC  27599-3255, 
USA.}

\maketitle
\date{\today}

\begin{abstract}
In an approach towards naturalness without supersymmetry,
the renormalization properties of 
nonsupersymmetric abelian quiver gauge theories are
studied. In the construction based on cyclic groups $Z_p$ the gauge 
group is
$U(N)^p$, the fermions are all in bifundamentals
and the construction allows scalars in adjoints and bifundamentals.
Only models without adjoint scalars, however, exhibit both chiral 
fermions
and the absence of one-loop quadratic divergences in the scalar
propagator.
\end{abstract}
\maketitle

\newpage

\noindent {\it Introduction}

One of the principal motivations for extending the standard model
of particle phenomenology, over the last three decades and more, has
been the concern about naturalness of the light Higgs scalar mass
({\it e.g.} \cite{JJ}).
One expects new physics to appear well above the weak scale,
for example at the Planck scale for gravity, at a GUT scale
for grand unification or at a right-handed neutrino
mass scale for the see-saw mechanism of neutrino mass. In
any such case the appearance of quadratic divergences in
the Higgs scalar propagator of the standard model would
suggest a heavy Higgs scalar mass thus necessitating fine-tuning
and unnaturalness.

There are several popular approaches to this naturalness question;
here we shall discuss a less popular but seemingly equally
valid approach. The popular ideas are (i) that the scalar is a bound
state analogous to a Cooper pair in BCS theory\cite{technicolor},
though no fully convincing model exists; (ii) that there
is an extra symmetry, supersymmetry\cite{susy}, which protects the 
light Higgs
mass once it is introduced by hand; (iii) that there
are large extra dimensions\cite{led} near the weak scale which avoid
the need of a much higher physics scale; (iv) that
among the $\sim 10^{100}$ vacua of string theory the
smallness $\sim 100$ GeV of the Higgs mass is correlated
to the smaller scale $\sim 1$ meV of cosmic dark 
energy\cite{landscape}.

An alternative approach is to use nonsupersymmetric gauge theories
derived from the most highly supersymmetric${\cal N} = 4$ gauge 
theories.
 Such nonsupersymmetric
${\cal N}=0$ gauge theories can be systematically constructed
\cite{F1998,FramptonVafa,F1999,F2003}
from ${\cal N}=4$ ones by using suitable orbifolding.
The resulting theories are not coupled to gravity; we here assume this
vanishing-gravity limit to be a sufficiently accurate approximation to 
the
physics of any foreseeable collider experiment, which would be 
sensitive
only to non-gravitational interactions.

Models can be constructed with four-dimensional conformal invariance
at high energies; for the models we consider here, the renormalization
group $\beta$-functions vanish for all SU(N) gauge groups.  However, 
other desirable properties require $U(N)$ gauge groups so
there is still a subtlety of decoupling U(1) factors (discussed later).

For example, in \cite{F1999} there is a $Z_7$ model which contains
all the states of the standard model and in \cite{F2003}
there is a $Z_{12}$ model allowing grand unification at a scale 4 TeV.
In the sequence of $Z_p$ models as we shall see
the first with chiral fermions is $Z_4$ but
the $Z_7$ and $Z_{12}$ examples also fall into the
class we shall investigate.
We shall then discuss the quadratic divergence of the
scalar propagator at one loop.

\bigskip
\bigskip

\noindent {\it Classification of abelian quiver gauge theories}

We consider the compactification of the type-IIB superstring
on the orbifold $AdS_5 \times S^5/\Gamma$
where $\Gamma$ is an abelian group $\Gamma = Z_p$
of order $p$ with elements ${\rm exp} \left( 2 \pi i A/p \right)$, 
$0 \le A \le (p-1)$.

The resultant quiver gauge theory has ${\cal N}$
residual supersymmetries with ${\cal N} = 2,1,0$ depending
on the details of the embedding of $\Gamma$
in the $SU(4)$ group which is the isotropy
of the $S^5$. This embedding is specified
by the four integers $A_m, 1 \le m \le 4$ with

\begin{equation}
\Sigma_m A_m = 0 {\rm (mod p)}
\label{SU4}
\end{equation}
which characterize
the transformation of the components of the defining
representation of $SU(4)$.

We are here interested in the non-supersymmetric
case ${\cal N} = 0$ which occurs if and only if
all four $A_m$ are non-vanishing.

The gauge group is $U(N)^p$. The fermions
are all in the bifundamental representations
\begin{equation}
\Sigma_{m=1}^{m=4}\Sigma_{j=1}^{j=p} (N_j, \bar{N}_{j + A_m})
\label{fermions}
\end{equation}
which are manifestly non-supersymmetric because no
fermions are in adjoint representations
of the gauge group.
Scalars appear in representations 
\begin{equation}
\Sigma_{i=1}^{i=3}\Sigma_{j=1}^{i=p} (N_j, \bar{N}_{j \pm a_i})
\label{scalars}
\end{equation}
in which the six integers $(a_i, -a_i)$ characterize the 
transformation of the
antisymmetric second-rank tensor representation 
of $SU(4)$. The $a_i$
are given by $a_1 = (A_2+A_3), a_2= (A_3+A_1), a_3= (A_1+A_2)$

It is possible for one or more of the $a_i$ to vanish
in which case the corresponding scalar representation
in the summation in Eq.(\ref{scalars}) is to be interpreted as an 
adjoint
representation of one particular $U(N)_j$.
One may therefore
have zero, two, four or all six of the scalar
representations, in Eq.(\ref{scalars}), in such adjoints.
It is one purpose of the present article to
investigate how the renormalization properties
and occurrence of quadratic divergences
depend on the distribution
of scalars into bifundamental
and adjoint representations.

Note that there is one model with all scalars in adjoints for each even
value of $p$ (see Model Nos 1,3,12). For general even $p$
the embedding is 
$A_m=(\frac{p}{2},\frac{p}{2},\frac{p}{2},\frac{p}{2})$. This series
is the complete list of ${\cal N}=0$ abelian quivers with
all scalars in adjoints.

To be of more phenomenolgical interest the model should
contain chiral fermions. This requires that the embedding
be complex: $A_m \not\equiv -A_m$ (mod p). It will now be shown
that for 
the presence of chiral fermions all scalars must be in bifundamentals.

The proof of this assertion follows by assuming the contrary,
that there is at least one adjoint arising from, say, $a_1=0$. 
Therefore
$A_3=-A_2$ (mod p). But then it follows from Eq.(\ref{SU4})
that $A_1=-A_4$ (mod p). The fundamental representation of $SU(4)$
is thus real and fermions are non-chiral\footnote{This is almost 
obvious
but for a complete justification, see\cite{FK03}}.

The converse also holds: If all $a_i \neq 0$ then there are chiral 
fermions.
This follows since by assumption
$A_1 \neq -A_2$, $A_1 \neq -A_3$, $A_1 \neq -A_4$. Therefore
reality of the fundamental representation would require
$A_1 \equiv -A_1$ hence, since $A_1 \neq 0$, $p$ is even 
and $A_1 \equiv \frac{p}{2}$; but then the other $A_m$
cannot combine to give only vector-like fermions.

It follows that:

\underline{{\it In an ${\cal N}=0$ quiver gauge theory, chiral fermions 
are possible}}

\underline{{\it if and only if all scalars are in bifundamental 
representaions.}}

\newpage

For the lowest few orders of the group $\Gamma$,
the members of the infinite
class of ${\cal N}=0$ abelian quiver gauge theories
are tabulated below:

\bigskip
\bigskip
\bigskip

\begin{tabular}{||c||c||c|c||c|c|c||c|c||}
\hline
Model No. & p & $A_m$ & $a_i$ & scalar & scalar & chiral & Contains \\
&&&& bifunds. & adjoints & fermions? & SM fields? \\
\hline
\hline
1 & 2 & (1111) & (000) & 0 & 6 & No  &  No \\
\hline
\hline
2 & 3 & (1122) & (001) & 2 & 4 & No  &  No \\
\hline
\hline
3 & 4 & (2222) & (000) & 0 & 6 & No  &  No \\
4 & 4 & (1133) & (002) & 2 & 4 & No  &  No \\
5 & 4 & (1223) & (011) & 4 & 2 & No  &  No \\
6 & 4 & (1111) & (222) & 6 & 0 & Yes  &  No \\
\hline
\hline
7 & 5 & (1144) & (002) & 2 & 4 & No  &  No \\
8 & 5 & (2233) & (001) & 2 & 4 & No  &  No \\
9 & 5 & (1234) & (012) & 4 & 2 & No  &  No \\
10 & 5 & (1112) & (222) & 6 & 0 & Yes  &  No \\
11 & 5 & (2224) & (111) & 6 & 0 & Yes  &  No \\
\hline
\hline
12 & 6 & (3333) & (000) & 0 & 6 & No  &  No \\
13 & 6 & (2244) & (002) & 2 & 4 & No  &  No \\
14 & 6 & (1155) & (002) & 2 & 4 & No  &  No \\
15 & 6 & (1245) & (013) & 4 & 2 & No  &  No \\
16 & 6 & (2334) & (011) & 4 & 2 & No  &  No \\
17 & 6 & (1113) & (222) & 6 & 0 & Yes  &  No \\
18 & 6 & (2235) & (112) & 6 & 0 & Yes  &  No \\
19 & 6 & (1122) & (233) & 6 & 0 & Yes  &  No \\
\hline
\hline
\end{tabular}

\newpage

\bigskip

The Table continues to infinity but we stop at $p=7$:

\bigskip
\bigskip
\bigskip
\begin{tabular}{||c||c||c|c||c|c|c||c|c||}
\hline
Model No. & p & $A_m$ & $a_i$ & scalar & scalar & chiral & Contains \\
&&&& bifunds. & adjoints & fermions? & SM fields? \\
\hline
\hline
20 & 7 & (1166) & (002) & 2 & 4 & No  &  No \\
21 & 7 & (3344) & (001) & 2 & 4 & No  &  No \\
22 & 7 & (1256) & (013) & 4 & 0 & No  &  No \\
23 & 7 & (1346) & (023) & 4 & 2 & No  &  No \\
24 & 7 & (1355) & (113) & 6 & 0 & No  &  No \\
25 & 7 & (1114) & (222) & 6 & 0 & Yes  &  No \\
26 & 7 & (1222) & (333) & 6 & 0 & Yes  &  No \\
27 & 7 & (2444) & (111) & 6 & 0 & Yes  &  No \\
28 & 7 & (1123) & (233) & 6 & 0 & Yes  &  Yes \\
29 & 7 & (1355) & (113) & 6 & 0 & Yes  &  Yes \\
30 & 7 & (1445) & (122) & 6 & 0 & Yes  &  Yes \\
\hline
\hline
\end{tabular}

\newpage

\bigskip
\bigskip
\bigskip
\bigskip

\noindent {\it Quadratic Divergences}

The lagrangian for the nonsupersymmetric $Z_p$ theory can be written in 
a convenient
notation which accommodates simultaneously both adjoint and 
bifundamental scalars as
\begin{eqnarray}
{\cal L} & = & 
-\frac{1}{4} F_{\mu\nu; r,r}^{ab}F_{\mu\nu; r,r}^{ba}
+i \bar{\lambda}_{r + A_4, r}^{ab} \gamma^{\mu} D_{\mu} \lambda_{r, 
r+A_4}^{ba}
+ 2 D_{\mu} \Phi_{r+a_i, r}^{ab \dagger} D_{\mu} \Phi_{r, r+a_i}^{ba}
+i \bar{\Psi}_{r+A_m, r}^{ab} \gamma^{\mu} D_{\mu} \Psi_{r, r+A_m}^{ba} 
\nonumber \\
&  & 
- 2 i g  
\left[ \bar{\Psi}_{r, r+A_i}^{ab} P_L \lambda_{r + A_i, r + A_i + 
A_4}^{bc}
\Phi_{r + A_i+A_4, r}^{\dagger ca}
- \bar{\Psi}_{r, r+A_i}^{ab} P_L \Phi_{r + A_i, r - A_4}^{\dagger bc} 
\lambda_{r - A_4, r}^{ca}
\right] \nonumber \\
& & -   \sqrt{2} i g \epsilon_{ijk}
\left[
\bar{\Psi}_{r, r + A_i}^{ab} P_L \Psi_{r +A_i, r + A_i + A_j}^{bc} 
\Phi_{r -A_k - A_4, r}^{ca}
-
\bar{\Psi}_{r, r + A_i}^{ab} P_L \Phi_{r +A_i, r + A_i + A_k + 
A_4}^{bc} \Psi_{r - A_j, r}^{ca}
\right] \nonumber \\
& & - g^2 \left(
\Phi_{r, r + a_i}^{ab} \Phi_{r+a_i,r}^{\dagger bc} 
- 
\Phi_{r, r - a_i}^{\dagger ab} \Phi_{r-a_i,r}^{bc} 
\right)
\left(
\Phi_{r, r + a_j}^{cd} \Phi_{r+a_j,r}^{\dagger da} 
-
\Phi_{r, r - a_j}^{\dagger cd} \Phi_{r- a_j,r}^{da} 
\right)  \nonumber \\
& & + 4 g^2
\left(
\Phi_{r, r+a_i}^{ab}\Phi_{r+a_i, r+a_i+a_j}^{bc}
\Phi_{r+a_i+a_j,r+a_j}^{\dagger cd}\Phi_{r+a_j,r}^{\dagger da}
- 
\Phi_{r, r+a_i}^{ab}\Phi_{r+a_i, r+a_i+a_j}^{bc}
\Phi_{r+a_i+a_j,r+a_i}^{\dagger cd}\Phi_{r+a_i, r}^{\dagger da}
\right)
\label{N=0L}
\end{eqnarray}
where $\mu, \nu = 0, 1, 2, 3$ are lorentz indices; $a, b, c, d = 1$ to 
$N$ are $U(N)^p$
group labels; $r = 1$ to $p$ labels the node of the quiver diagram
(when the two node subscripts are equal it is an adjoint plus singlet
and the two superscripts are in the same U(N): when the two node
subscripts are unequal it is a bifundamental and the two superscript
labels transform under different U(N) groups);
$a_i ~~ (i = \{1, 2, 3\}) $ label the first three of the {\bf 6} of 
SU(4);
$A_m ~~ (m = \{1, 2, 3 ,4\}) = (A_i, A_4)$ label the {\bf 4} of SU(4). 
By definition
$A_4$ denotes an arbitrarily-chosen fermion ($\lambda$) associated with 
the gauge boson,
similarly to the notation in the ${\cal N} = 1$ supersymmetric case.
Recall that $\sum_{m=1}^{m=4} A_m = 0 $ (mod p).

\bigskip

As we showed in the previous section, the infinite sequence of 
nonsupersymmetric
$Z_p$ models can have scalars in adjoints (corresponding to $a_i = 0$)
and bifundamentals ($a_i \neq 0$). Denoting by $x$ the
number of the three $a_i$ which are non-zero, the models with $x=3$
have only bifundamental scalars, those with $x=0$ have only
adjoints while $x=1,2$ models contain both types of scalar 
representations.
As we have seen, to contain the phenomenologically-desirable chiral 
fermions, it
is necessary and sufficient that $x=3$.

Let us first consider the quadratic divergence question in the mother
${\cal N} = 4$ theory. The ${\cal N}=4$ lagrangian is like 
Eq.(\ref{N=0L})
but since there is only one node all those subscripts become 
unnecessary
so the form is simply
\begin{eqnarray}
{\cal L} & = & 
-\frac{1}{4} F_{\mu\nu}^{ab}F_{\mu\nu}^{ba}
+i \bar{\lambda}^{ab} \gamma^{\mu} D_{\mu} \lambda^{ba}
+ 2 D_{\mu} \Phi_{i}^{ab \dagger} D_{\mu} \Phi_{i}^{ba}
+i \bar{\Psi}_{m}^{ab} \gamma^{\mu} D_{\mu} \Psi_{m}^{ba} \nonumber \\
&  & 
- 2 i g 
\left[\bar{\Psi}_{i}^{ab} P_L \lambda^{bc}
\Phi_{i, r}^{\dagger ca}
- \bar{\Psi}_{i}^{ab} P_L \Phi_{i}^{bc}\lambda^{ca}
\right] \nonumber \\
& & - \sqrt{2} i g \epsilon_{ijk}
\left[
\bar{\Psi}_{i}^{ab} P_L \Psi_{j}^{bc} \Phi_{k}^{\dagger ca}
-
\bar{\Psi}_{i}^{ab} P_L \Phi_{j}^{bc} \Psi_{k}^{ca}
\right] \nonumber \\
& & - g^2 \left(
\Phi_{i}^{ab} \Phi_{i}^{\dagger bc} 
- 
\Phi_{i}^{\dagger ab} \Phi_{i}^{bc} 
\right)
\left(
\Phi_{j}^{cd} \Phi_{j}^{\dagger da} 
-
\Phi_{j}^{\dagger cd} \Phi_{j}^{da} 
\right)  \nonumber \\
& & + 4 g^2
\left(
\Phi_{i}^{ab}\Phi_{j}^{bc}
\Phi_{i}^{\dagger cd}\Phi_{j}^{\dagger da}
- 
\Phi_{i}^{ab}\Phi_{j}^{bc}
\Phi_{j}^{\dagger cd}\Phi_{i}^{\dagger da}
\right)
\label{N=4L}
\end{eqnarray}

All ${\cal N} = 4$ scalars are in adjoints and the scalar propagator
has one-loop quadratic divergences coming potentially from
three scalar self-energy diagrams:
(a) the gauge loop (one quartic vertex);
(b) the fermion loop (two trilinear vertices);
and (c) the scalar loop (one quartic vertex).

For ${\cal N} = 4$ the respective contributions
of (a, b, c) are computable from
Eq.(\ref{N=4L}) as proportional to $g^2N (1, -4, 3)$ which cancel
exactly.

The ${\cal N} = 0$ results for the scalar self-energies (a, b, c)
are computable from the lagrangian of Eq.(\ref{N=0L}).
Fortunately, the calculation was already done in \cite{BFLP}.
The result is amazing! The quadratic divergences cancel
if and only if x = 3, exactly the same ``if and only if" 
as to have chiral fermions.
It is pleasing that one can independently confirm
the results of \cite{BFLP} directly from the interactions
in Eq.(\ref{N=0L}) To give just one explicit
example, in the contributions to
diagram (c) from the last term in Eq.(\ref{N=0L}), the
1/N corrections arise from a contraction of $\Phi$ with 
$\Phi^{\dagger}$
when all the four color superscripts are distinct
and there is consequently no sum
over color in the loop. For this case, examination of the node 
subscripts then
confirms proportionality to the kronecker delta, $\delta_{0, a_i}$.
If and only if all $a_i \neq 0$, all the other terms
in Eq.(\ref{N=0L}) do not lead to 1/N corrections
to the ${\cal N}=4$. 

Some comments on the literature are necessary. In the 1999
paper of Csaki, Skiba and Terning\cite{CST} it was claimed that
there are always $1/N$ corrections to spoil cancellation
for finite N  and that $N \geq 10^{28}$ is necessary! This
was because of a technical error that the orbifolded
gauge group is not $SU(N)^p$ but $U(N)^p$ and bifundamentals
carry U(1) charges. A paper by Fuchs\cite{fuchs} in 2000,
which has been largely ignored, corrected this point.

The conclusion is that the chiral $Z_7, Z_{12}$ models of 
\cite{F1999,F2003}
which contain the standard model
are free of one-loop quadratic divergences in the scalar propagator.
Nevertheless the overall conformal invariance would not be respected
by U(1) factors which would have non-zero positive beta-functions.
Clearly these factors must somehow be decoupled.
This mysterious decoupling of U(1)'s  from AdS/CFT which 
would not be conformally invariant has been commented 
upon in \cite{witten,polchinski}.
A better understanding of these U(1)'s may be necessary to
achieve the hope of a fully four-dimensionally
conformally invariant extension of the standard model.
There is the paradoxical requirement that the U(1)
gauge factors must be present to cancel quadratic
divergences but must decouple to preserve 4-dim
conformal invariance at high energy.

Eventually gravity, at the Planck scale, will
inevitably break conformal invariance because
Newton's constant is dimensionful. A realistic
hope is that there is a substantial window of
energy scales where conformal invariance 
is an excellent approximation between, say,
4 TeV \cite{F2003} for at least a few orders of magnitude
in energy even towards a scale approaching the see-saw
scale $\sim 10^{10}$ GeV. It is difficult to foresee
how large the conformality window is.
Finally it is interesting to note that the 
present models seem to have all the ingredients of the 
so-called little Higgs models\cite{littlehiggs},
which were proposed later than \cite{F1998}, with the
quiver diagram here interpreted as the theory space there.

\bigskip
\bigskip

\section*{Acknowledgements}

One of us (PHF) thanks D.R.T. Jones for discussions. 
This work was supported in part by the
U.S. Department of Energy under Grant
No. DE-FG02-97ER-41036


\bigskip
\bigskip

\newpage

\bigskip
\bigskip


\begin{thebibliography}{99}
\bibitem{JJ}
I. Jack and D.R.T. Jones, Phys. Lett. {\bf B234,} 321 (1990).
\bibitem{technicolor}
S. Weinberg, Phys. Rev. {\bf D13,} 974 (1975); {\it ibid} {\bf D19,} 
1277 (1979).
L. Susskind, Phys. Rev. {\bf D20,} 2619-2625 (1979); E. Farhi and L. 
Susskind, Phys. Reports, {\bf 74,} 277 (1981).
\bibitem{susy}
J. Wess and B. Zumino, Nucl. Phys. {\bf B70,} 39-50 (1974);
H.P. Nilles, Phys. Reports, {\bf 110,} 1 (1984). 
H.E. Haber and G.L. Kane, Phys. Reports, {\bf 117,} 75 (1985).
\bibitem{led}
I. Antoniadis, Phys. Lett. {\bf B246,} 377-384 (1990).
N. Arkani-Hamed, S. Dimopoulos and G.R. Dvali,
Phys. Lett. {\bf B429,} 263-272 (1998). {\tt hep-ph/9803315}.
I. Antoniadis, N. Arkani-Hamed, S. Dimopoulos and G.R. Dvali,
Phys. Lett. {\bf B436,} 257-263 (1998). {\tt hep-ph/9804398}.
\bibitem{landscape}
W. Lerche, D. Lust and A.N. Schellekens, Nucl. Phys. {\bf B287,} 477 
(1987).\\
R. Bousso and J. Polchinski, JHEP 06: 006 (2000).
\bibitem{F1998}
P.H. Frampton, Phys. Rev. {\bf D60,} 085004 (1999). {\tt 
hep-th/9812117}.
\bibitem{FramptonVafa}
P.H. Frampton and C. Vafa. {\tt hep-th/9903226}
\bibitem{F1999}
P.H. Frampton, Phys. Rev. {\bf D60,} 121901 (1999). {\tt 
hep-th/9907051}. see also: \\
{\it ibid} 085004 (1999). {\tt hep-th/9905042}.
\bibitem{F2003}
P.H. Frampton, Mod. Phys. Lett. {\bf A18,} 1377 (2003). {\tt 
hep-ph/0208044}. see also: \\
P.H. Frampton, R.M. Rohm and T. Takahashi, Phys. Lett. {\bf B570,} 67 
(2003).
{\tt hep-ph/0302074}.
\bibitem{FK03}
P.H. Frampton and T.W. Kephart, Int. J. Mod. Phys. {\bf A19,} 593 
(2004).
{\tt hep-th/0308207}.
\bibitem{BFLP}
P. Brax, A. Falkowski, Z. Lalak and S. Pokorski, Phys. Lett. {\bf 
B538,} 426-434 (2002).
\bibitem{CST}
C. Csaki, W. Skiba and J. Terning, Phys. Rev. {\bf D61,} 025019 (1999).
\\
In this paper, the orbifolded gauge group was assumed to be $SU(N)^p$
rather than $U(N)^p$ and hence bifundamentals
which do have U(1) charges were incorrectly incorporated.
\bibitem{fuchs}
E. Fuchs, JHEP 0010: 028 (2000). {\tt hep-th/0003235}.
\bibitem{witten}
E. Witten, Adv. Theor. Math. Phys. {\bf 2,} 253-291 (1998).
{\tt hep-ph/9802150}. Footnote 7. 
\bibitem{polchinski}
J. Polchinski, Int. J. Math. Phys. {\bf A16,} 707-718 (2001).
{\tt hep-th/0011193}. Footnote at end of Section 3.1.
\bibitem{littlehiggs}
N. Arkani-Hamed, A.G. Cohen, T. Gregoire, E. Katz,
A.E. Nelson and J.G. Wacker, JHEP 0208:021 (2002).
{\tt hep-ph/0206020}.
\end{thebibliography}
\end{document}